\documentclass[aps,prd,showpacs,eqsecnum,twocolumn]{revtex4}
\usepackage{amsmath,amssymb}
\usepackage{graphicx}

\begin{document}

\title{Numerical experiments of adjusted BSSN systems for controlling
constraint violations}

\author{Kenta Kiuchi$^{1}$
\footnote{\affiliation\ kiuchi@gravity.phys.waseda.ac.jp}}
\author{Hisa-aki Shinkai$^{2}$
\footnote{\affiliation\ shinkai@is.oit.ac.jp}}
\affiliation{$^{1}$Department of Physics, Waseda University, 3-4-1 Okubo,
 Shinjuku-ku, Tokyo 169-8555, Japan~}
\affiliation{$^{2}$Faculty of Information Science and Technology, Osaka Institute of Technology, 1-79-1 Kitayama, Hirakata, Osaka 573-0196, Japan~ }

\date{\today}

\begin{abstract}
We present our numerical comparisons between the BSSN formulation 
widely used in numerical relativity today
and its adjusted versions using constraints.  
We performed three testbeds: gauge-wave, linear wave, and 
Gowdy-wave tests, proposed by the Mexico workshop on the 
formulation problem of the Einstein equations. 
We tried three kinds of adjustments, which were previously proposed 
from the analysis of the constraint propagation equations, and investigated
how they improve the accuracy and stability of evolutions. 
We observed that the signature of the proposed Lagrange multipliers are
always right and the adjustments improve the convergence
and stability of the simulations.  When the original BSSN system
already shows satisfactory good evolutions (e.g., linear wave test), 
the adjusted versions also coincide with those evolutions; 
while in some cases (e.g., gauge-wave or Gowdy-wave tests) 
the simulations using the adjusted systems last 10 times as long as 
those using the original BSSN equations.
Our demonstrations imply a potential to construct a robust evolution 
system against constraint violations 
even in highly dynamical situations. 
\end{abstract}

\pacs{04.20.-q, 04.25.Dm, 04.25.-g}

\maketitle

%\baselineskip 18pt
%*********************
\section{Introduction}
%*********************

Numerical integration of the Einstein equations is the only way to investigate
highly dynamical and nonlinear gravitational space-time. 
The detection of gravitational wave requires templates of waveform, 
among them mergers of compact objects are the most plausible astrophysical 
sources.  Numerical relativity has been developed with this purpose
over decades. 

For neutron star (NS) binaries, a number of scientific numerical 
simulations have been done so far, and we are now at the level of 
discussing the actual physics of the phenomena, including the effects of 
the equations of state, hydrodynamics, and general relativity 
by evolving various initial data~
\cite{Shibata:2003ga,Shibata:2006nm,Marronetti:2003gk,Marronetti:2003hx,Faber:2006qc}. 
Mergers of black holes (BHs) are also available after the breakthrough by 
Pretorius~\cite{Pretorius:2005gq} in 2004. 
Pretorius's implementation had many novel features in his code;
among them he 
discretizes the four-dimensional Einstein equations directly, 
which is not a conventional approach so far.  
However, after the announcements of successful  
binary BH mergers by Campanelli et al.~\cite{Campanelli:2005dd} and 
Baker et al.~\cite{Baker2006} based on the standard 3+1 decomposition of the Einstein equations, 
many groups began producing interesting results~\cite{Diener2006,Herrmann2006,Baker:2006yw,Gonzalez:2006md,Etienne:2007hr,Tichy:2007hk,Campanelli:2007cg,
Gonzalez:2007hi,Campanelli:2007ew,Thornburg:2007hu}. 
The merger of NS-BH binary simulations has also been reported recently, 
e.g. ~\cite{Shibata:2006bs}.

Almost all the groups which apply the above conventional approach use
the so-called BSSN variables together with ``$1+\log$''-type slicing conditions
for the lapse function and ``$\Gamma$-driver'' type slicing conditions
for the shift function.  BSSN stands for Baumgarte-Shapiro~\cite{BS} and 
Shibata-Nakamura~\cite{SN}, the modified Arnowitt-Deser-Misner formulation
initially proposed by Nakamura~\cite{Nakamura}.  
(The details are described in \S \ref{subsec:BSSN}.)
There have already been 
several efforts to explain why the combination of this recipe works 
from the point of view of the well-posedness of the partial differential equations
(e.g.~\cite{Beyer2004,Gundlach2006}). However, 
the question remains whether there exists an alternative evolution system 
that enables 
 more long-term stable and accurate simulations. 
The search for a better set of equations for numerical integrations is called
the formulation problem for numerical relativity, of which earlier stages are
reviewed by one of the authors~\cite{novabook}. 

In this article, we report our numerical tests of modified versions 
of the BSSN system, the {\it adjusted BSSN systems}, 
proposed by Yoneda and Shinkai~\cite{Yoneda:2002kg}. 
The idea of their modifications is to add constraints to the evolution 
equations like Lagrange multipliers and to construct a robust evolution 
system which evolves to the constraint surface as the attractor. 
Their proposals are based on the eigenvalue analysis of the constraint 
propagation equations (the evolution equations of the constraints) 
on the perturbed metric.  For the ADM formulation,  they explain
why the standard ADM does not work for long-term simulations by
showing the existence of the constraint violating mode 
in perturbed Schwarzschild space-time \cite{Shinkai2002}. 
For the BSSN formulation, they analyzed the eigenvalues 
of the constraint propagation equations only on flat space-time
~\cite{Yoneda:2002kg}, but 
one of their proposed adjustments was immediately tested 
by Yo et al.~\cite{YoBaumgarteShapiro} for the numerical evolution of 
Kerr-Schild space-time and 
confirmed to work as expected.
(The details are described in \S \ref{subsec:adjBSSN}.)

Our numerical examples are taken from the proposed problems for 
testing the formulations of the Mexico Numerical 
Relativity Workshop 2001 participants~\cite{Alcubierre:2003pc}, 
which are sometimes called the Apples-with-Apples test. 
To concentrate the comparisons on the formulation problem, 
the templated problems are settled so as not to require technical complications; 
e.g., 
periodic boundary conditions are used and the slicing conditions 
do not require solving elliptical equations. 
Several groups already reported their code tests
using these Apples tests
(e.g. \cite{Jansen:2003uh,Zlochower:2005bj,Boyl2007}), 
and we are also able to compare our results with theirs. 

This article is organized as follows. 
We describe the BSSN equations and the {\it adjusted} BSSN equations 
in Sec.~\ref{subsec:BSSN} and \ref{subsec:adjBSSN}. 
We give our three numerical test problems 
in Sec.~\ref{sec:setup}. 
Comments on our coding stuff are in Sec.~\ref{sec:code}. 
Sec.~\ref{sec:num} is devoted to showing numerical results for each testbeds, 
and we summarize the results in 
Sec.~\ref{sec:summary}.

%*****************************************
\section{Basic equations}\label{sec:basic}
%*****************************************
%*********************************************
\subsection{BSSN equations}\label{subsec:BSSN}
%*********************************************
We start by presenting the standard BSSN formulation, 
where we follow the notations of~\cite{BS}, which are widely used among numerical relativists.

The idea of the BSSN formulation is to introduce auxiliary variables  
to those of the Arnowitt-Deser-Misner (ADM) formulation for obtaining longer 
stable numerical simulations. 
The basic variables of the BSSN formulation are 
$(\phi,\tilde{\gamma}_{ij},K,
\tilde{A}_{ij},\tilde{\Gamma}^i)$, which are defined by 
%---------------------------------------------------------------------------
\begin{eqnarray}
\phi &=& \frac{1}{12} \log(\text{det} \gamma_{ij}),\label{eq:conf}\\
\tilde{\gamma}_{ij} &=& e^{-4\phi} \gamma_{ij},\label{eq:con-metric}\\
K &=& \gamma^{ij}K_{ij},\label{eq:ext-tr}\\
\tilde{A}_{ij} &=& e^{-4\phi} \left[ K_{ij} - \frac{1}{3}\gamma_{ij}K 
\right],\label{eq:ext-trldess}\\
\tilde{\Gamma}^i &=& \tilde{\gamma}^{jk}{\tilde{\Gamma}^i}_{jk},
\label{eq:conf-con}
\end{eqnarray}
where $(\gamma_{ij}, K_{ij})$ are the intrinsic and extrinsic ADM 3-metric. 
The conformal factor $\phi$ is introduced so as to set  
$\tilde{\gamma} \equiv \text{det} [\tilde{\gamma}_{ij}]$ as unity, 
$\tilde{A}_{ij}$ is supposed to be traceless, and
$\tilde{\Gamma}^i$ is treated independently in evolution equations.  Therefore
these three requirements turn into the new constraints 
[below (\ref{eq:cons-calG})-(\ref{eq:cons-calS})]. 

The set of the BSSN evolution equations are 
\begin{widetext} 
%----------------------------------------------------------------------
\begin{eqnarray}
\partial_t \phi &=& - \frac{1}{6} \alpha K + \beta^i 
\partial_i \phi + \frac{1}{6} \partial_i \beta^i, \label{eq:ev-conf}\\
\partial_t \tilde{\gamma}_{ij} &=& - 2 \alpha \tilde{A}_{ij} 
+ \tilde{\gamma}_{ik} \partial_j \beta^k 
+ \tilde{\gamma}_{jk} \partial_i \beta^k
- \frac{2}{3}\tilde{\gamma}_{ij} \partial_k \beta^k 
+ \beta^k \partial_k \tilde{\gamma}_{ij},\label{eq:ev-con-metric}\\
\partial_t K &=& - D^i D_i \alpha 
+ \alpha \tilde{A}_{ij} \tilde{A}^{ij} + \frac{1}{3}\alpha K^2
+ \beta^i \partial_i K,\label{eq:ev-ext-tr}\\
\partial_t \tilde{A}_{ij} &=& - e^{-4\phi}\left[ D_i D_j \alpha 
+ \alpha R_{ij} \right]^{\text{TF}} 
+ \alpha K \tilde{A}_{ij}
- 2 \alpha \tilde{A}_{ik} {\tilde{A}^k}_{~j} \nonumber \\
&&+ \partial_i \beta^k \tilde{A}_{kj}
+ \partial_j \beta^k \tilde{A}_{ki}
- \frac{2}{3} \partial_k \beta^k \tilde{A}_{ij} 
+ \beta^k \partial_k \tilde{A}_{ij},\label{eq:ev-ext-trless}\\
\partial_t \tilde{\Gamma}^i &=& - 2 \partial_j \alpha \tilde{A}^{ij} 
+ 2 \alpha \left[ {\tilde{\Gamma}^i}_{jk} \tilde{A}^{jk} 
- \frac{2}{3}\tilde{\gamma}^{ij}\partial_j K 
+ 6 \tilde{A}^{ij} \partial_j \phi \right] \nonumber \\
&&+ \tilde{\gamma}^{jk}\partial_j\partial_k\beta^i
+ \frac{1}{3}\tilde{\gamma}^{ij}\partial_j\partial_k \beta^k
+ \beta^j \partial_j \tilde{\Gamma}^i
- \tilde{\Gamma}^j \partial_j \beta^i
+ \frac{2}{3}\tilde{\Gamma}^i \partial_j \beta^j \label{eq:ev-conf-con},
\end{eqnarray}
%----------------------------------------------------------------------
where $D_i$ is the covariant derivative with respect to the 3-metric 
$\gamma_{ij}$ and TF means trace-free operation, i.e., $H_{ij}^{\text{TF}} 
= H_{ij}-
\frac{1}{3}\gamma_{ij}{H^k}_k$. 
The Ricci tensor is computed with the 
conformal connection $\tilde{\Gamma}^i$ as 
%-----------------------------------------------------------------------
\begin{eqnarray}
&&R_{ij} = R^\phi_{ij} + \tilde{R}_{ij},\label{eq:ricci}\\
&&R^\phi_{ij} = - 2 \tilde{D}_i\tilde{D}_j \phi - 2 \tilde{\gamma}_{ij}
\tilde{D}^k \tilde{D}_k \phi + 4 \tilde{D}_i \phi \tilde{D}_j \phi
- 4 \tilde{\gamma}_{ij} \tilde{D}^k \phi \tilde{D}_k \phi 
\label{eq:cof-ricci}\\
&&\tilde{R}_{ij} = - \frac{1}{2}\tilde{\gamma}^{lk}\partial_k \partial_l 
\tilde{\gamma}_{ij} 
+ \tilde{\gamma}_{k(i} \partial_{j)}\tilde{\Gamma}^k
+ \tilde{\gamma}^{lm}\tilde{\Gamma}^k_{lm} \tilde{\Gamma}_{(ij)k}
+ 2 \tilde{\gamma}^{lm} \tilde{\Gamma}^k_{l(i} \tilde{\Gamma}_{j)km}
+ \tilde{\gamma}^{lm} \tilde{\Gamma}^k_{im}\tilde{\Gamma}_{klj},
\label{eq:conm-ricci}
\end{eqnarray}
%------------------------------------------------------------------------
\end{widetext} 
where $\tilde{D}_i$ is a covariant derivative associated with 
$\tilde{\gamma}_{ij}$.

Similarly to the ADM formulation, this system has constraint equations. 
The two ``kinematic" constraints, the Hamiltonian and momentum constraint equations, 
are expressed in terms of the BSSN basic variables and are written as  
\begin{widetext} 
%------------------------------------------------------------------------
\begin{eqnarray}
&&{\cal H} = e^{-4\phi}\tilde{R} - 8 e^{-4\phi} ( \tilde{D}^i \tilde{D}_i 
\phi + \tilde{D}^i \phi \tilde{D}_i \phi ) 
+ \frac{2}{3} K^2 
- \tilde{A}_{ij} \tilde{A}^{ij} - \frac{2}{3} {\cal A} K \approx 0,\label{eq:cal H}\\
&&{\cal M}_i = 6 {\tilde{A}^j}_{~i} \tilde{D}_j \phi 
- 2 {\cal A} \tilde{D}_i \phi - \frac{2}{3} \tilde{D}_i K
+ \tilde{D}_j {\tilde{A}^j}_{~i}\approx 0.\label{eq:cal Mi}
\end{eqnarray}
%------------------------------------------------------------------------
\end{widetext} 
Additionally, the BSSN formulation requires three ``algebraic" constraint relations; 
%----------------------------------------------------------------------
\begin{eqnarray}
{\cal G}^i &=& \tilde{\Gamma}^i - \tilde{\gamma}^{jk}
{\tilde{\Gamma}^i}_{jk}\approx 0,\label{eq:cons-calG}\\
{\cal A} &=& \tilde{A}_{ij} \tilde{\gamma}^{ij}\approx 0,\label{eq:cons-calA}\\
{\cal S} &=& \tilde{\gamma} - 1\approx 0, \label{eq:cons-calS}
\end{eqnarray}
%----------------------------------------------------------------------
where (\ref{eq:cons-calG}) and (\ref{eq:cons-calA}) are from the definitions of 
(\ref{eq:conf-con}) and (\ref{eq:ext-trldess}), respectively.  
Equation~(\ref{eq:cons-calS})
is from the requirement on $\tilde{\gamma}$. 

These five constraints are, theoretically, supposed to be zero at all times; 
therefore they can be used to modify the dynamical equations. For example, 
Alcubierre et al.~\cite{Alcubierre2000}
announced that the replacement of the terms in (\ref{eq:ev-conf-con})
using the momentum constraint drastically changes the stability feature. 
Actually, such replacements of terms using constraints are 
applied (with/without intentions) in many terms in 
(\ref{eq:ev-conf})-(\ref{eq:ev-conf-con}), which are expressed as 
Eqs.~(2.27)-(2.31) in~\cite{Yoneda:2002kg}. 

Alcubierre et al.~\cite{Alcubierre:2000xu} also pointed out 
that the re-definition of $\tilde{A}_{ij}$ by 
\begin{eqnarray}
\tilde{A}_{ij} \to \tilde{A}_{ij} - \frac{1}{3}\tilde{\gamma}_{ij} 
\text{tr} 
\tilde{A} \label{redef_traceout}
\end{eqnarray}
during the time evolutions improves the numerical stability. 
This technique again can be understood as the trace-out of the 
${\cal A}$-constraint (\ref{eq:cons-calA})  
from the evolution equations. 
In our numerical code, we do not apply this technique 
because we recognize the trace-free property as the 
new constraint ${\cal A}$ in the BSSN system, and 
our purpose is to 
construct a system preventing the violation of constraints.

Recently, several groups applied artificial dissipation 
(e.g.~\cite{Kreiss:1973}) to obtain stable evolutions 
(see, e.g.~\cite{Babiuc:2007vr,Zlochower:2005bj,Bruegmann:2006at}). 
We, however, do not introduce such dissipations in our code, 
since we try to clarify the difference of stability from the viewpoint of
 {\it formulations} of the Einstein equations.

%*******************************************************
\subsection{Adjusted BSSN systems}\label{subsec:adjBSSN}
%*******************************************************

To understand the stability property of the BSSN system, 
Yoneda and Shinkai~\cite{Yoneda:2002kg} studied 
the structure of the evolution equations, 
(\ref{eq:ev-conf})-(\ref{eq:ev-conf-con}), in detail, especially
how the modifications using the constraints, 
(\ref{eq:cal H})-(\ref{eq:cons-calS}), affect to the stability. 
They investigated the signature of the eigenvalues of the constraint 
propagation equations (dynamical equations of constraints), 
and explained that the standard BSSN dynamical equations are balanced 
from the viewpoints of constrained propagations, including a clarification 
of the 
effect of the replacement using the momentum constraint equation. 

Moreover, they 
predicted that several combinations of modifications have a 
constraint-damping nature, and named them {\it adjusted} BSSN systems. 
(Their predictions are based on the signature of eigenvalues of the 
constraint propagations, and the negative signature implies 
a dynamical system which evolves toward the constraint surface 
as the attractor.) 

Among them, in this work, we test the following three adjustments: 
\begin{enumerate}
\item An adjustment of the $\tilde{A}$-equation with the 
momentum constraint:  
\begin{eqnarray}
\partial_t \tilde{A}_{ij} = \partial^B_t \tilde{A}_{ij} 
+ \kappa_{A} \alpha \tilde{D}_{(i} {\cal M}_{j)}, \label{B1-adj}
\end{eqnarray}
where $\kappa_{\cal A}$ is predicted (from the eigenvalue analysis)
to be positive in order to 
damp the constraint violations. 

\item An adjustment of the $\tilde{\gamma}$-equation 
with ${\cal G}$ constraint:
\begin{eqnarray}
\partial_t \tilde{\gamma}_{ij} = \partial_t^B \tilde{\gamma}_{ij}
+ 
\kappa_{\tilde{\gamma}} \alpha \tilde{\gamma}_{k(i} 
\tilde{D}_{j)} {\cal G}^k, \label{B2a-adj}
\end{eqnarray}
with $\kappa_{\tilde{\gamma}} < 0 $.

\item An adjustment of the $\tilde{\Gamma}$-equation 
with ${\cal G}$ constraint:
\begin{eqnarray}
\partial_t \tilde{\Gamma}^i = \partial_t^B \tilde{\Gamma}^i 
+ 
\kappa_{\tilde{\Gamma}} \alpha {\cal G}^i.
\label{B2b-adj}
\end{eqnarray}
with $\kappa_{\tilde{\Gamma}} < 0$.
\end{enumerate}

These three adjustments are chosen
as samples of ``best candidates",
Eq.~(4.9)-(4.11) in~\cite{Yoneda:2002kg}.
The term ``best" comes from their conjecture on the eigenvalue analysis
of the constraint propagation matrix; that is, (a) all the resultant 
eigenvalues from above adjustments can be less than or at most equal 
to zero, which indicates the
decay of constraint errors, and (b) the resultant constraint propagation matrix
is diagonalizable, which guarantees the predictions of above eigenvalue
analysis (see Table II in~\cite{Yoneda:2002kg}).  
However, since above eigenvalues include zero elements and also above 
analysis assumes %casts on 
a linearly perturbed metric about the flat space-time, 
the effects of the adjustments (\ref{B1-adj})-(\ref{B2b-adj}) 
need to be demonstrated 
via numerical experiments.

%**************************************************
\section{Numerical Testbed Models}\label{sec:setup}
%**************************************************

Following the proposals of the Mexico Numerical Relativity Workshop~
\cite{Alcubierre:2003pc}, we perform three kinds of tests. 
In this section, we explicitly give some details of the models. 

%***************************************************
\subsection{Gauge-wave testbed}\label{sec:Gaugewave}
%***************************************************
The first test is the trivial Minkowski space-time, but sliced with 
the time-dependent 3-metric, which is called the gauge-wave test. 
The 4-metric is obtained by coordinate transformation from the 
Minkowski metric as 
%--------------------------------------------------------------
\begin{eqnarray}
ds^2 = - H dt^2 + H dx^2 + dy^2 + dz^2, \label{eq:gwave-metric}
\end{eqnarray}
%--------------------------------------------------------------
where
%------------------------------------------------------------
\begin{eqnarray}
H = H(x-t) = 1 - A \sin \left( \frac{2 \pi (x-t)}{d} \right), 
\label{eq:gwave-func}
\end{eqnarray}
%------------------------------------------------------------
which describes a sinusoidal gauge wave of amplitude $A$ propagating 
along the $x$-axis.  The non-trivial extrinsic curvature is 
%-------------------------------------------------------------------------
\begin{eqnarray}
K_{xx} = - \frac{\pi A}{d} \frac{\cos \left( \frac{2 \pi (x-t)}{d}\right)}
{\sqrt{ 1 + A \sin \frac{2 \pi (x-t)}{d} } }. \label{eq:gwave-extric}
\end{eqnarray}
%-------------------------------------------------------------------------
Following \cite{Alcubierre:2003pc}, we chose 
numerical domain and parameters as follows:
\begin{itemize}
\item Gauge-wave parameters: $d = 1$ and $A=10^{-2}$
\item Simulation domain: $x {\cal 2} [-0.5,0.5]$, $y=z=0$
\item Grid: $x^i = - 0.5 + ( n - \frac{1}{2} )dx$ with $n=1,\cdots 50 \rho$, 
where 
$dx=1/(50\rho)$ with $\rho = 2,4,8$
\item Time step: $dt=0.25 dx$
\item Boundary conditions:
Periodic boundary condition in $x$ direction and planar symmetry in 
$y$ and $z$ directions
\item  Gauge conditions:
%---------------------------------------------------------------------
\begin{eqnarray}
\partial_t \alpha = - \alpha^2 K,~~~\beta^i = 0.\label{eq:gwave-gauge}
\end{eqnarray}
%---------------------------------------------------------------------
\end{itemize}
The 1D simulation is carried out for a $T=1000$ crossing-time or until the 
code crashes, where one crossing-time is defined by the length of 
the simulation domain.

%*****************************************************
\subsection{Linear wave testbed}\label{sec:Linearwave}
%*****************************************************
The second test is to check the ability of handling a travelling gravitational wave.  
The initial 3-metric and extrinsic curvature $K_{ij}$ are given by a 
diagonal perturbation with component
%-------------------------------------------------------------------------
\begin{eqnarray}
ds^2 = - dt^2 + dx^2 + ( 1 + b )dy^2 + ( 1 - b )dz^2, \label{eq:lin-metric}
\end{eqnarray}
%--------------------------------------------------------------------------
where
%-----------------------------------------------------------------
\begin{eqnarray}
b = A \sin \left( \frac{2\pi (x-t)}{d} \right) \label{eq:lin-func}
\end{eqnarray}
%-----------------------------------------------------------------
for a linearized plane wave traveling in the $x$-direction. Here $d$ is 
the linear size of the propagation domain and $A$ is the amplitude of 
the wave. The non-trivial components of extrinsic curvature are then
%----------------------------------------------------------------------
\begin{eqnarray}
K_{yy} = - \frac{1}{2} \partial_t b,~~K_{zz} = \frac{1}{2}\partial_t b.
\end{eqnarray}
%----------------------------------------------------------------------
Following \cite{Alcubierre:2003pc}, we chose the following parameters:
\begin{itemize}
\item Linear wave parameters: $d = 1$ and $A=10^{-8}$
\item Simulation domain: $x {\cal 2} [-0.5,0.5]$, $y=0$, $z=0$
\item Grid: $x^i = - 0.5 + ( n - \frac{1}{2} )dx$ with $n=1,\cdots 50 \rho$, where
$dx=1/(50\rho)$ with $\rho = 2,4,8$
\item Time step: $dt=0.25 dx$
\item Boundary conditions:
Periodic boundary condition in $x$ direction and planar symmetry in 
$y$ and $z$ directions
\item Gauge conditions: $\alpha=1$ and $\beta^i=0$
\end{itemize}
The 1D simulation is carried out for a $T=1000$ crossing-time or until the 
code crashes. 

%******************************************************
\subsection{Collapsing polarized Gowdy-wave testbed}\label{sec:Gowdywave}
%******************************************************
The third test is to check the formulation in a strong field context using
the polarized Gowdy metric, which is written as 
%-------------------------------------------------------------------
\begin{eqnarray}
ds^2 = t^{-1/2}e^{\lambda/2}(-dt^2+dz^2)+t(e^P dx^2 + e^{-P} dy^2 ).
\label{eq:Gowdy-metric}
\end{eqnarray}
%------------------------------------------------------------------
Here time coordinate $t$ is chosen such that time increases as the universe 
expands. Simple forms of the solutions, $P$ and $\lambda$, are given by 
%------------------------------------------------------------------
\begin{eqnarray}
 P &=& J_0 (2 \pi t) \cos(2\pi z),\label{eq:Gowdy-funcP}\\
 \lambda &=& - 2\pi t J_0 ( 2\pi t ) J_1 ( 2\pi t)\cos^2(2\pi z) 
\nonumber\\
&& + 
2\pi^2 t^2[J_0^2(2\pi t) + J_1^2(2\pi t)]
\nonumber\\
&& - \frac{1}{2}[(2\pi)^2 [J_0^2(2\pi) 
+ J_1^2(2\pi)] \nonumber \\
&&- 2\pi J_0(2\pi) J_1(2\pi)],\label{eq:Gowdy-funcl}
\end{eqnarray}
%-------------------------------------------------------------------
where $J_n$ is the Bessel function. The non-trivial extrinsic curvatures 
are then
%----------------------------------------------------------------------------
\begin{eqnarray}
&&K_{xx} = - \frac{1}{2} t^{1/4} e^{-\lambda/4} e^P ( 1 + t P_{,t} ),\\
\label{eq:Gowdy-extxx}
&&K_{yy} = - \frac{1}{2} t^{1/4} e^{-\lambda/4} e^{-P} ( 1 - t P_{,t} ),\\
\label{eq:Gowdy-extyy}
&&K_{zz} = \frac{1}{4}t^{-1/4} e^{\lambda/4} ( t^{-1} - \lambda_{,t} ).
\label{eq:Gowdy-extzz}
\end{eqnarray}
%---------------------------------------------------------------------------
According to~\cite{Alcubierre:2003pc}, the new time coordinate $\tau$, 
which satisfies 
harmonic condition, is obtained by coordinate transformation as 
%----------------------------------------
\begin{eqnarray}
t(\tau) = k e^{c\tau},\label{eq:new-time}
\end{eqnarray}
%----------------------------------------
where $c$ and $k$ are arbitrary constants. Using this freedom, we can set 
the lapse function in the new time coordinate to be unity at the initial time. 
Concretely, we set
%---------------------------------------------------------------
\begin{eqnarray}
t_0 = \tau_0 &\sim& 9.8753205829098, \nonumber \\
c &\sim& 0.0021195119214617, \\
k &\sim& 9.6707698127638, \nonumber 
\end{eqnarray}
%---------------------------------------------------------------
where $t_0$ is the initial time. Following~\cite{Alcubierre:2003pc}, 
we perform our evolution in the collapsing (i.e. backward in time) direction. 
Parameters are chosen as follows:
\begin{itemize}
\item Simulation domain: $z$ ${\cal 2}$ $[-0.5,0.5]$, $x=y=0$
\item Grid: $z = - 0.5 + ( n - \frac{1}{2} )dz$ with $n=1,\cdots 50 \rho$, where
$dz=1/(50\rho)$ with $\rho = 2,4,8$
\item Time step: $dt=0.25 dz$
\item Boundary conditions: Periodic boundary condition in $z$-direction 
and plane symmetry in $x$- and $y$-directions
\item Gauge conditions: the harmonic slicing~(\ref{eq:gwave-gauge}) 
and $\beta^i=0$
\end{itemize}
The 1D simulation is carried out for a $T=1000$ crossing-time or until the 
code crashes. 

%*********************************
\section{The Code}\label{sec:code}
%*********************************
%****************************
\subsection{Code description}
%****************************
We have developed a new numerical code based on the adjusted BSSN systems.
The variables are $(\phi, \tilde{\gamma}_{ij}, K,
\tilde{A}_{ij}, \tilde{\Gamma}^i)$, and the evolution equations are
(\ref{eq:ev-conf})-(\ref{eq:ev-conf-con}) with/without adjustment
(\ref{B1-adj}), (\ref{B2a-adj}), and/or (\ref{B2b-adj}).
The time-integration is under the free-evolution scheme, and we monitor
five constraints, (\ref{eq:cal H})-(\ref{eq:cons-calS}), to check the
accuracy and stability of the evolutions. 

Our time-integration scheme is the three-step iterative Crank-Nicholson method with centered finite difference in space~\cite{Teukolsky:1999rm}.
This scheme should have second-order convergence both in space and time, and we checked its convergence in all the testbeds. 

As we have already mentioned in the end of \S II A, we do not apply the 
trace-out technique of $\tilde{A}_{ij}$, (\ref{redef_traceout}) in our
code.

We also remark on our treatment of the conformal connection 
variable $\tilde{\Gamma}^i$.  As was pointed out in~\cite{Alcubierre:2002kk}, it is better not to use 
$\tilde{\Gamma}^i$ in all the evolution equations. 
We surmise this is because the amplification of the error due to 
the discrepancy of the definition (\ref{eq:conf-con}), i.e., 
the accumulations of the violations of ${\cal G}^i$-constraint 
(\ref{eq:cons-calG}).  Therefore, we used the evolved $\tilde{\Gamma}^i$
only for the terms in (\ref{eq:ev-conf-con}) and (\ref{eq:conm-ricci}), 
and not for other terms, so as not to implicitly apply the 
${\cal G}^i$-constraint in time evolutions. 

%************************************************
\subsection{Debugging procedures}\label{subsec:debug}
%***********************************************
It is crucial that our code can produce accurate results, 
because the adjustment 
methods are based on the assumption that the code represents 
the BSSN system~(\ref{eq:ev-conf})-(\ref{eq:ev-conf-con}) accurately. 
We verified our code 
by comparing our numerical 
data with analytic solutions from %like 
the gauge-wave and Gowdy-wave testbeds
in Sec.~\ref{sec:setup}.  
The actual procedures are as follows:
\begin{enumerate}
\item Evolve only one component, e.g. $\tilde{A}_{xx}$, numerically, and
express all the other components with those of the analytic solution. 
In this situation, the origin of the error is from the 
finite differencing of the analytic solution in the spatial direction and 
from that of the numerically evolved 
component ($\tilde{A}_{xx}$) both in spatial and time directions. 
We checked the code by monitoring the difference between 
the numerically evolved component ($\tilde{A}_{xx}$) and its 
analytic expression.  This procedure was applied to 
all the components one by one.
\item Evolve only several components, 
e.g., $\tilde{A}_{xx}$ and $\tilde{\Gamma}^x$, 
numerically, and express the other components by the analytic solution. 
The error can be checked by a procedure similar to the one above. 
\item Evolve all the components numerically, and check the error
with the analytic solution. 
\end{enumerate}

We repeated these procedure three times by switching the 
propagation directions ($x$, $y$, and $z$-directions) 
of gauge-wave and Gowdy-wave solutions. 
We also applied these procedures in a 2D test~\cite{Alcubierre:2003pc}, and 
checked the off-diagonal component.  

%***********************************
\subsection{Error evaluation methods}
%***********************************
It should be emphasized that the adjustment effect has two meanings, improvement of stability and of accuracy. 
Even if a simulation is stable, 
it does not imply that the result is accurate. 
We judge the stability of the evolution by monitoring the 
L2 norm of each constraint, 
%---------------------------------------------------------------
\begin{eqnarray}
||\delta {\cal C}||_2 (t) \equiv \sqrt{\frac{1}{N}
\sum_{x,y,z} \left( {\cal C}(t; x,y,z) \right)^2}, \label{eq:errorC}
\end{eqnarray}
where $N$ is the total number of grid points, 
%----------------------------------------------------------------
while we judge the accuracy by 
the difference of the metric components $g_{ij}(t; x,y,z)$ 
from the exact solution $g_{ij}^{\text{(exact)}}
(t; x,y,z)$, 
%---------------------------------------------------------------
\begin{eqnarray}
||\delta g_{ij}||_2 (t) \equiv \sqrt{\frac{1}{N}
\sum_{x,y,z} \left( g_{ij}  - g_{ij}^{\text{(exact)}}
  \right)^2}.\label{eq:error}
\end{eqnarray}
%----------------------------------------------------------------

%*********************************
\subsection{Magnitude of $\kappa$}
%*********************************
Adjusted systems,~(\ref{B1-adj})-(\ref{B2b-adj}), require to specify 
the parameter $\kappa$. 
From the analytical prediction in ~\cite{Yoneda:2002kg} 
we know the signature of $\kappa$, but not for its magnitude.
By definition of the adjustment terms in Eq.~(\ref{B1-adj})-(\ref{B2b-adj}), 
applying small $\kappa$ should produce the close results with those of the 
plain system. 
On the contrary, the large $\kappa$ system will violate the 
Courant-Friedrich-Lewy condition~\cite{hyp2}.
Hence, there exists a suitable region in the adjustment 
parameters.

At this moment, we have to chose $\kappa$ experimentally, by observing 
the life-time of simulations.  The value of $\kappa$, used in our 
demonstrations, is one of the choices of  
which the adjustment works effectively in all the 
resolutions. 

%****************************************
\section{Numerical Results}\label{sec:num}
%****************************************
%***************************
\subsection{Gauge-wave test}
%***************************
%********************************
\subsubsection{The plain BSSN system}
\begin{figure*}
  \begin{center}
  \vspace*{40pt}
    \begin{tabular}{cc}
      \resizebox{90mm}{!}{\includegraphics{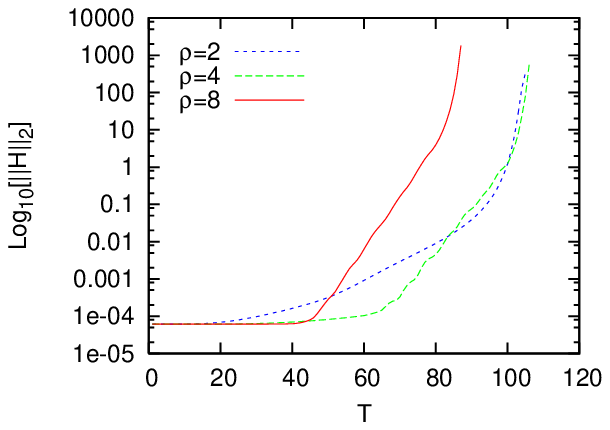}} &
      \resizebox{90mm}{!}{\includegraphics{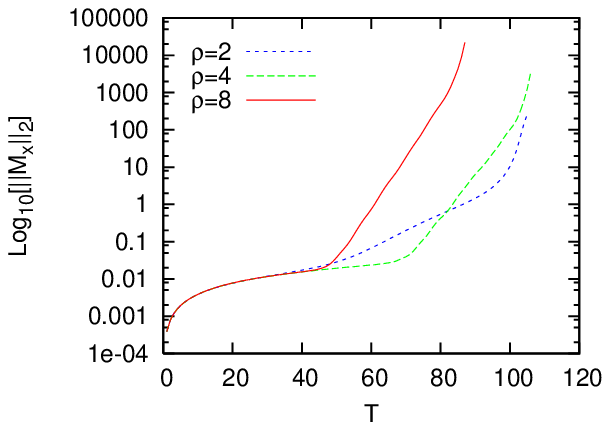}}\\
    \end{tabular}
    \caption{\label{1D-Gauge-plain}
     The one-dimensional gauge-wave test with the plain BSSN system. 
     The L2 norm of ${\cal H}$ and ${\cal M}_x$, rescaled by $\rho^2/4$, are plotted 
     with a function of the crossing-time. 
     The amplitude of the wave is $A=0.01$.  
     The loss of convergence at the early time, near the 20 
     crossing-time, can be seen, 
     and it will produce the blow-ups of the calculation in the end.}
  \end{center}
\end{figure*}

%********************************
As the first test, we show the plain BSSN evolution (that is, no adjustments) in 
Fig.~\ref{1D-Gauge-plain} for the gauge-wave test. In Fig.~\ref{1D-Gauge-plain}, 
the L2 norms of the Hamiltonian and momentum 
constraints (\ref{eq:errorC}) are plotted as a 
function of the crossing-time. 
The second-order convergent nature is lost at an early time, the 
20 crossing-time, and the simulation crashes at about the 100 
crossing-time. 
The poor performance of the plain BSSN system for the gauge wave test 
has been reported in~\cite{Jansen:2003uh} (see their Fig.~8). 
This drawback, on the other hand, can be overcome
if one uses the fourth-order finite differencing scheme, 
an example of which can be seen in~\cite{Zlochower:2005bj} (see their Fig.~2).

%-----------------------------------------------------------------------------
%-----------------------------------------------------------------------------

%******************************************************
\subsubsection{Adjusted BSSN with $\tilde{A}$-equation}
\begin{figure*}
  \begin{center}
  \vspace*{40pt}
    \begin{tabular}{cc}
      \resizebox{90mm}{!}{\includegraphics{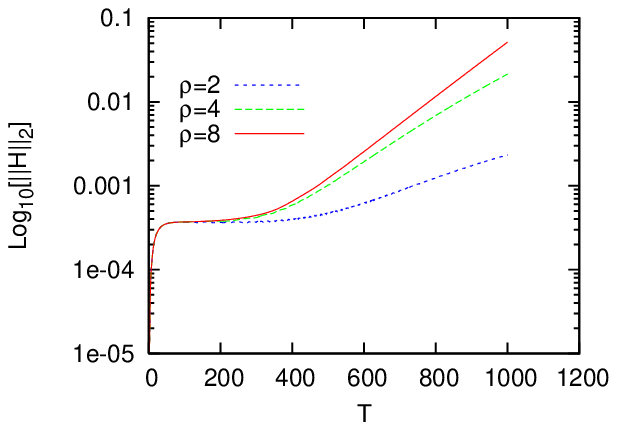}} &
      \resizebox{90mm}{!}{\includegraphics{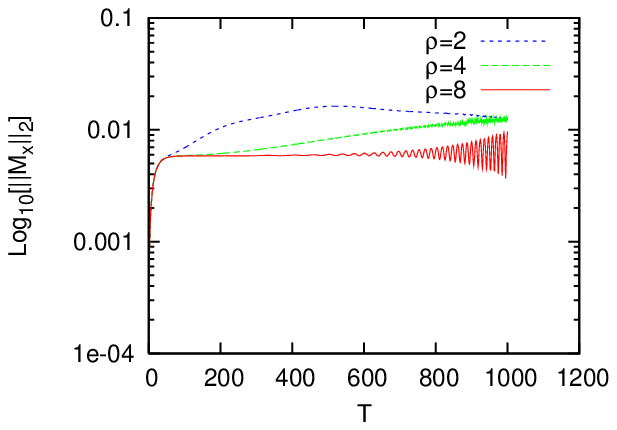}}\\
    \end{tabular}
    \caption{\label{1D-Gauge-B1}
     The one-dimensional gauge-wave test with the adjusted BSSN system in 
     the $\tilde{A}$-equation (\ref{B1-adj}). 
     The L2 norm of ${\cal H}$ and ${\cal M}_x$, rescaled by $\rho^2/4$, are plotted 
     with a function of the crossing-time. 
     The wave parameter is the same as with Fig.~\ref{1D-Gauge-plain}, and the 
     adjustment parameter 
     $\kappa_{A}$ is set to $\kappa_{A}=0.005$. 
     We see the higher resolution runs show convergence longer, i.e., 
     the 300 crossing-time in ${\cal H}$ and the 200 crossing-time 
     in ${\cal M}_x$ with $\rho=4$ and $8$ runs.
     All runs can stably evolve up to the 1000 crossing-time.}
  \end{center}
\end{figure*}

%******************************************************
We found that the simulation lasts 
10 times longer with the adjustment in the $\tilde{A}$-equation using the momentum constraint (\ref{B1-adj}). 
Figure~\ref{1D-Gauge-B1} shows the L2 norms of the Hamiltonian and Momentum 
constraints in the same style as in Fig.~\ref{1D-Gauge-plain}. 
The adjustment parameter is set at $\kappa_{A}=0.005$ for this plot.  
We obtain almost prefect overlap of the rescaled Hamiltonian 
constraint for 200 crossing-times and almost perfect overlap 
in the momentum constraint for 50 crossing-times;
there apparently improve the results of the plain BSSN system
(see Fig.~\ref{1D-Gauge-plain}).
We show the plots until the 1000 crossing-time, there we observe
the growth of the error both in later time and in higher resolution
cases.  However, it is also true that all errors are still under the
errors of the plain BSSN system. 
Therefore, we conclude that this adjusted system shows a 
weaker instability than the plain system.

%-----------------------------------------------------------------------------
%-----------------------------------------------------------------------------

%*************************************************************
\subsubsection{Adjusted BSSN with $\tilde{\Gamma}$-equation}
\begin{figure*}
  \begin{center}
  \vspace*{40pt}
    \begin{tabular}{cc}
      \resizebox{90mm}{!}{\includegraphics{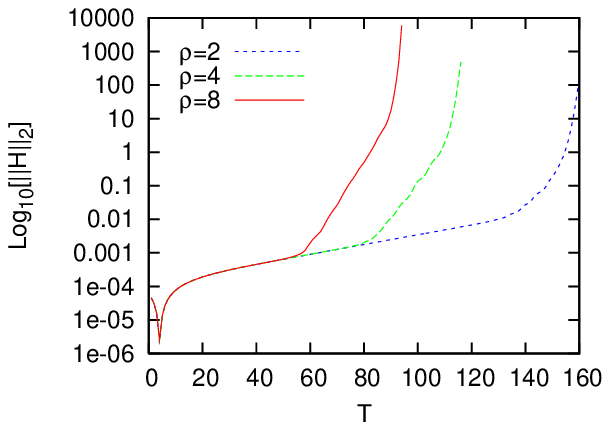}} &
      \resizebox{90mm}{!}{\includegraphics{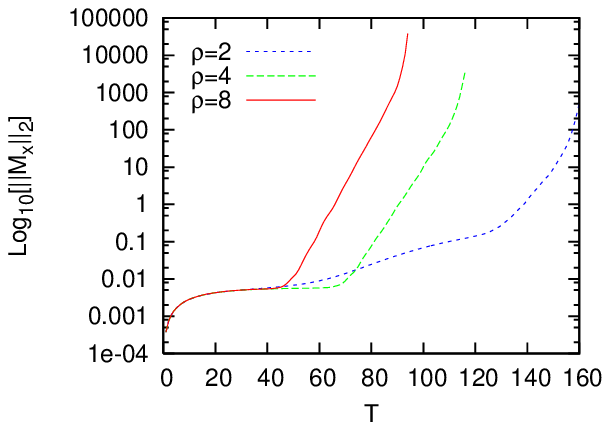}}\\
    \end{tabular}
    \caption{\label{1D-Gauge-B2b}
     The one-dimensional gauge-wave test with the adjusted BSSN system in 
     the $\tilde{\Gamma}$-equation (\ref{B2b-adj}). 
     The L2 norm of ${\cal H}$ and ${\cal M}_x$, rescaled by $\rho^2/4$, are plotted 
     with a function of the crossing-time. 
     The wave parameter is the same as Fig.~\ref{1D-Gauge-plain}, and the 
     adjustment parameter is $\kappa_{\tilde{\Gamma}}=-0.1$. 
     Note the near perfect overlap for the 55 crossing-time in ${\cal H}$ 
     and the 40 crossing-time in ${\cal M}_x$.}
  \end{center}
\end{figure*}

%*************************************************************
The case of the adjustment of the $\tilde{\Gamma}$-equation 
using the ${\cal G}$ constraint (\ref{B2b-adj}) is shown in 
Fig.~\ref{1D-Gauge-B2b}. 

The adjustment parameter is set at $\kappa_{\tilde{\Gamma}}=-0.1$. 
We find that the second-order convergence breaks down near the 40 
crossing-time under the Momentum constraint, which is almost the same as with 
the plain BSSN system. However, the convergence of the Hamiltonian 
constraint is improved, i.e., it continues to the near 55 crossing-time. 
The life-time of the simulation is almost the same as that of 
the plain BSSN system. 
%-----------------------------------------------------------------------------
%-----------------------------------------------------------------------------

%*************************************************************
\subsubsection{Adjusted BSSN with $\tilde{\gamma}$-equation}
%*************************************************************
We also tested the cases of the adjustment of the 
$\tilde{\gamma}$-equation using the ${\cal G}$ constraint,  
(\ref{B2a-adj}).  
We again observed the effects of the adjustment on 
its stability and accuracy but found a rather 
small effect compared to the cases of 
the adjustments of (\ref{B1-adj}) or (\ref{B2b-adj}), up to our 
trials of the parameter range of $\kappa_{\gamma}$. 
Therefore we omit showing the results.

%*************************************
\subsubsection{Evaluation of Accuracy}
\begin{figure*}
  \begin{center}
    \begin{tabular}{cc}
      \resizebox{90mm}{!}{\includegraphics{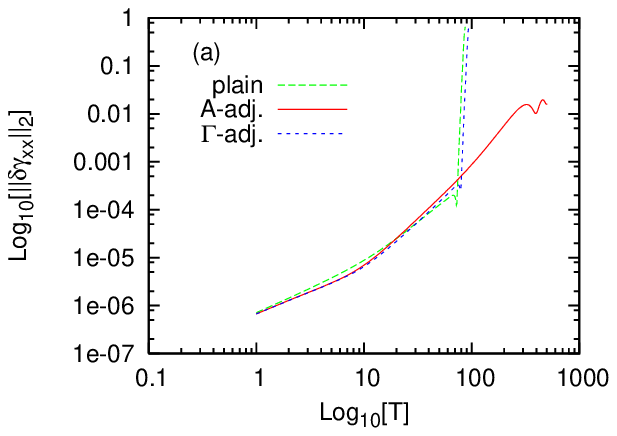}} &
      \resizebox{90mm}{!}{\includegraphics{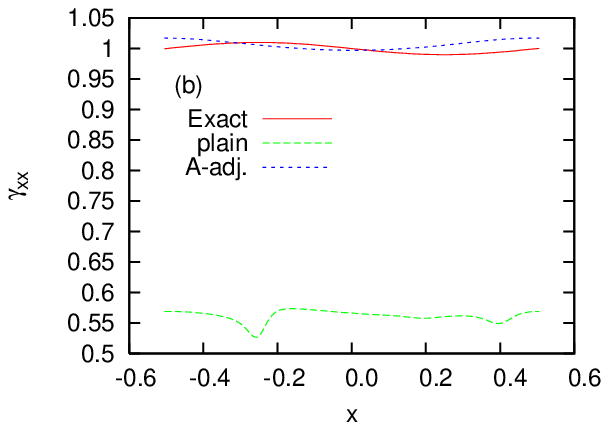}}\\
    \end{tabular}
    \caption{Evaluation of the accuracy of the one-dimensional gauge-wave testbed.  
     Lines show the plain BSSN, the adjusted BSSN with ${\cal A}$-equation, 
     and with ${\tilde{\Gamma}}$-equation.  
     (a) The L2 norm of the error in $\gamma_{xx}$, 
    using (\ref{eq:error}). 
	 (b) A snapshot of the exact and numerical solution at $T=100$.
     \label{1D-Gauge-gxx}}
  \end{center}
\end{figure*}

%*************************************
For evaluating the accuracy, we prepare Fig.~\ref{1D-Gauge-gxx}(a), in 
which we plot 
the L2 norm of the error in $\gamma_{xx}$, (\ref{eq:error}), 
with the function of time. 
Three lines correspond to the result of the plain BSSN system, 
$\tilde{A}$-eq. adjusted, and $\tilde{\Gamma}$-eq. adjusted 
BSSN system, respectively. 
The $\tilde{\Gamma}$-adjustment makes the life-time slightly longer than 
that of the plain BSSN, while $\tilde{A}$-adjustment 
increases the life-time of the simulation by a factor of 10.
However, it is also true that the error 
grows in time in all the three cases. 

We also find that the error is induced by distortion of the wave, i.e. 
the both phase and amplitude errors distort the numerical solution.  
In Fig.~\ref{1D-Gauge-gxx}(b), we show a snapshot 
of $\gamma_{xx}$ numerical solution at $T=100$, 
together with the exact solution at the same time coordinate. 
The amplitude difference between the numerical and exact solutions is 
apparently less when we use the $\tilde{A}$-eq. adjusted system 
than that of the plain system. 
In Sec.~\ref{sec:summary} later, we discuss 
what causes the error and why the simulation life-time becomes longer 
when we use the adjusted system.

%-----------------------------------------------------------------------------
%-----------------------------------------------------------------------------

%**********************************************
\subsection{Linear wave test}\label{sec:Linear}
\begin{figure*}
 \begin{center}
  \includegraphics[width=90mm]{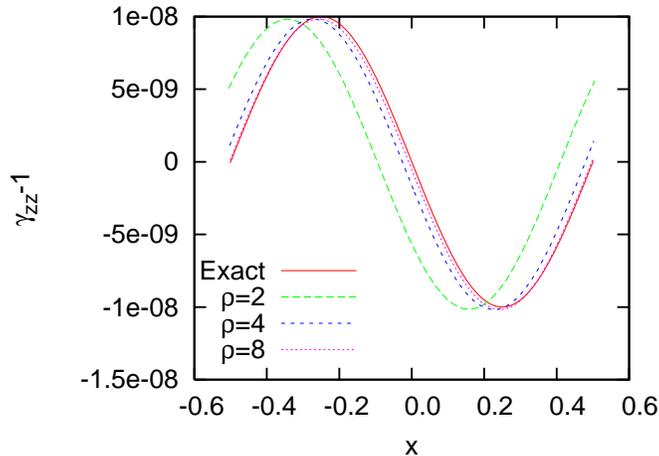}
   \caption{\label{1D-Linear} Snapshots of the one-dimensional linear wave at different resolutions with the plain BSSN system at the simulation time 
   500 crossing-time.  We see there exists phase error, but they are 
convergent away at higher resolution runs.}
 \end{center}
\end{figure*}

\begin{figure*}
 \begin{center}
  \includegraphics[width=90mm]{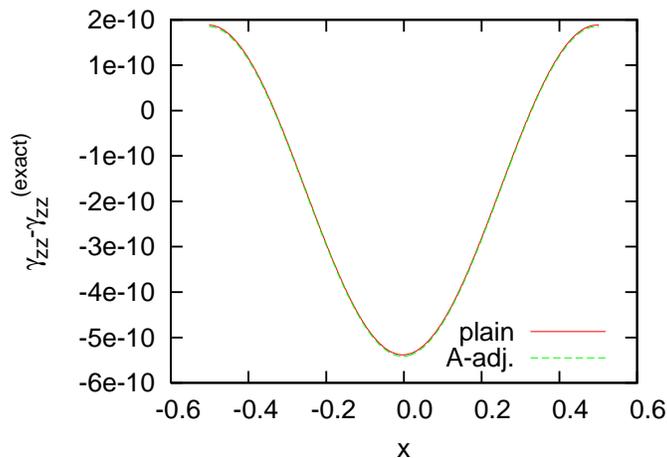}
   \caption{\label{1D-Linear-Phase} Snapshot of errors with the 
exact solution 
for the Linear Wave testbed with the plain BSSN system and 
the adjusted BSSN system with the $\tilde{A}$ equation at $T=500$. 
The highest resolution $\rho=8$ is used in both runs. 
The difference between the plain and the adjusted BSSN system with the 
$\tilde{A}$ equation is indistinguishable. Note that the maximum amplitude 
is set to be $10^{-8}$ in this problem.}
 \end{center}
\end{figure*}

%**********************************************
The second test is the linear wave propagation test, \S III B, 
to check the accuracy of wave propagations in the adjusted systems.
We find that the linear wave testbed does not produce 
a significant constraint violation even for the plain BSSN system. 
The simulation does not crash at the 1000 crossing-time irrespective of the 
resolutions.  Figure~\ref{1D-Linear} illustrates the profiles of 
$\gamma_{zz}-1$ at the 500 crossing-time.  The figure indicates the 
simulation does not produce the amplitude error but does produce 
the phase error. 
However, we also observe that the higher resolution run reduces the phase error.
%-----------------------------------------------------------------------
%-----------------------------------------------------------------------
We tried the same evolutions with adjusted BSSN systems.  However, 
all the results are indistinguishable from the those of the plain BSSN system. 
This is because the adjusted terms of the equations are small due to the small violations of constraints. 
Figure~\ref{1D-Linear-Phase} shows a snapshot of the error defined by 
$\gamma_{zz}-\gamma_{zz}^{\text{(exact)}}$ 
at the 500 crossing-time 
both for the plain BSSN system and 
the adjusted BSSN system where the $\tilde{A}$-equation 
where $\kappa_{\cal A}=10^{-3}$. 
Since two lines are matching quite well, we can say 
that the adjusted BSSN system produces the same 
result as the plain BSSN system, including the phase error. 
Results from the other adjusted BSSN systems 
are almost the same qualitatively, including their convergence features. 
We also remark that we do not see a case in which 
adjustment worsens accuracy and stability. 

%-----------------------------------------------------------------------
%-----------------------------------------------------------------------

%***********************************************
\subsection{Gowdy-wave test}\label{subsec:Gowdy}
%***********************************************
The third test is the polarized Gowdy-wave test, \S III C, to check the 
adjustments in the strong field regime. 
%******************************************************
\subsubsection{The plain BSSN}\label{subsubsec:plain-Gowdy}
\begin{figure*}
  \begin{center}
  \vspace*{40pt}
    \begin{tabular}{cc}
      \resizebox{90mm}{!}{\includegraphics{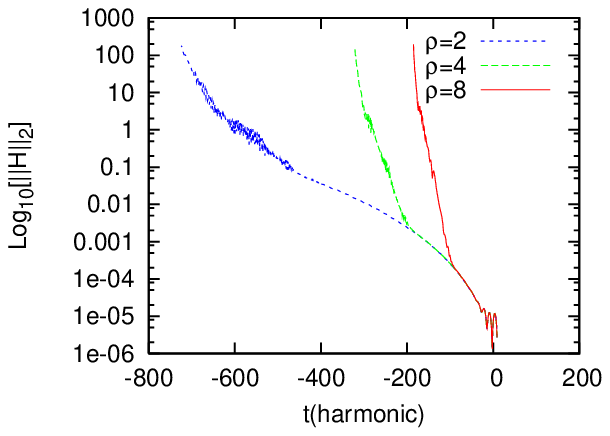}} &
      \resizebox{90mm}{!}{\includegraphics{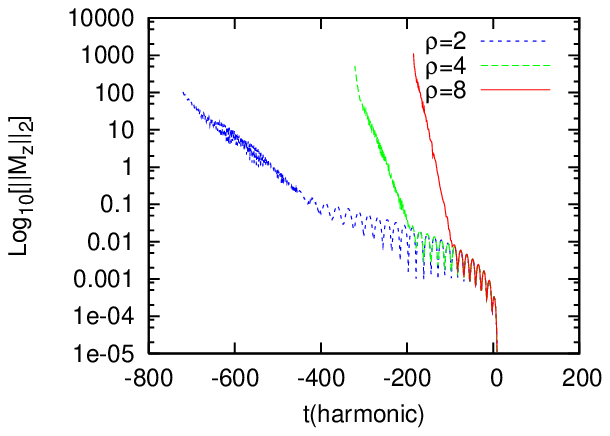}}\\
    \end{tabular}
    \caption{\label{1D-Gowdy-plain}
    Collapsing polarized Gowdy-wave test with the plain BSSN system. 
     The L2 norm of ${\cal H}$ and ${\cal M}_z$, rescaled by $\rho^2/4$,
     are plotted with a function of the crossing-time. 
    (Simulation proceeds backwards from $t=0$.) 
     We see almost perfect overlap for the initial 100 crossing-time, 
and the higher 
     resolution runs crash earlier.  This result is 
quite similar to 
those achieved with the 
Cactus BSSN code, reported by~\cite{Alcubierre:2003pc}.}
  \end{center}
\end{figure*}

%******************************************************
In Fig.~\ref{1D-Gowdy-plain}, 
We first show the case of the plain BSSN evolution. 
We find that the 
second-order convergence continues up-to the 100 crossing-time and 
the higher resolutions runs tend to crash at early times.
This behavior (and crashing time) almost coincides with the results
of the {\it Cactus} BSSN code, reported by Alcubierre et al.~\cite{Alcubierre:2003pc}
(see their Fig.~7). 
(We remark that 
Zlochower et al.~\cite{Zlochower:2005bj} reported 
they can produce the stable 
and accurate evolution for the 1000 crossing-time by implementing the
higher order differencing scheme to their {\it LazEv} code. 
However, it should be emphasized that they suggested 
their code produces the stable simulation only when they used the 
Kreiss-Oliger dissipation term~\cite{Kreiss:1973}.
)
%-----------------------------------------------------------------------------
%-----------------------------------------------------------------------------
%******************************************************
\subsubsection{Adjusted BSSN with $\tilde{A}$-equation}
\label{subsubsec:A-Gowdy}
\begin{figure*}
  \begin{center}
  \vspace*{40pt}
    \begin{tabular}{cc}
      \resizebox{90mm}{!}{\includegraphics{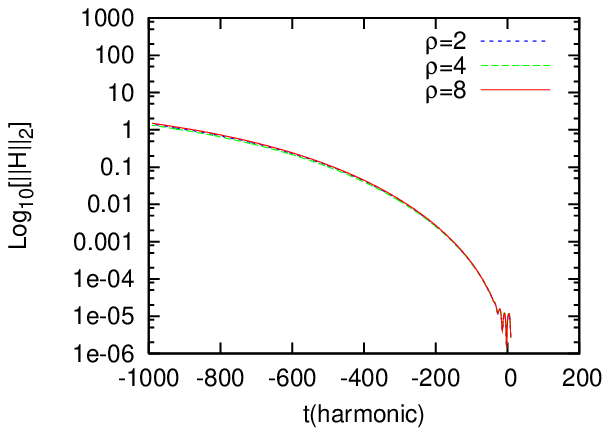}} &
      \resizebox{90mm}{!}{\includegraphics{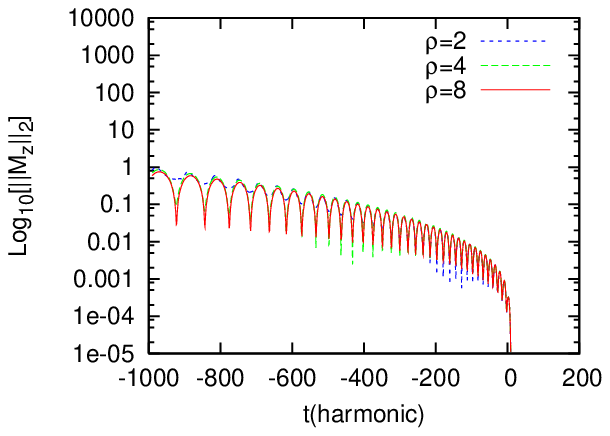}}\\
    \end{tabular}
    \caption{\label{1D-Gowdy-B1} 
     Collapsing polarized Gowdy-wave test with the adjusted BSSN system in 
     the $\tilde{A}$-equation (\ref{B1-adj}),  with $\kappa_{{\cal A}}=-0.001$.
     The style is the same as in Fig.~\ref{1D-Gowdy-plain} 
     and note that both constraints are normalized by $\rho^2/4$.
     We see almost perfect overlap for the initial 1000 crossing-time in both 
     constraint equations, 
     ${\cal H}$ and ${\cal M}_z$, even for the
     highest resolution run. }
  \end{center}
\end{figure*}

%******************************************************
Adjustment of the $\tilde{A}$-equation using the momentum constraint (\ref{B1-adj}), extends the life-time of the simulation 10 times longer for the highest resolution run.  
Figure~\ref{1D-Gowdy-B1} depicts the 
rescaled 
L2 norm of ${\cal H}$ and ${\cal M}_z$ 
versus time.  We set $\kappa_{{\cal A}}=-0.001$. (Note that 
the signature of $\kappa$ is reversed from the expected one, since 
the evolution is backward in time.)

We find that an almost perfect overlap up to the 
1000 crossing-time under both  
the Hamiltonian constraint and the Momentum constraint. 
(These overlaps indicate that the error in ${\cal H}$ and ${\cal M}_z$
in the $\rho=8$ resolution runs are 16 times smaller than these errors in the
$\rho=2$ resolution run. )
However, we also find oscillations 
in the Momentum constraint, especially in the end of the simulation.
%-----------------------------------------------------------------------------
%-----------------------------------------------------------------------------
%**************************************************************
\subsubsection{Adjusted BSSN with $\tilde{\gamma}$-equation}
\label{subsubsec:g-Gowdy}
\begin{figure*}
  \begin{center}
  \vspace*{40pt}
    \begin{tabular}{cc}
      \resizebox{90mm}{!}{\includegraphics{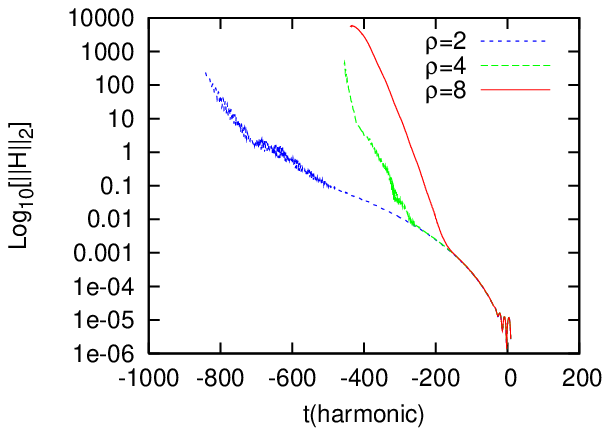}} &
      \resizebox{90mm}{!}{\includegraphics{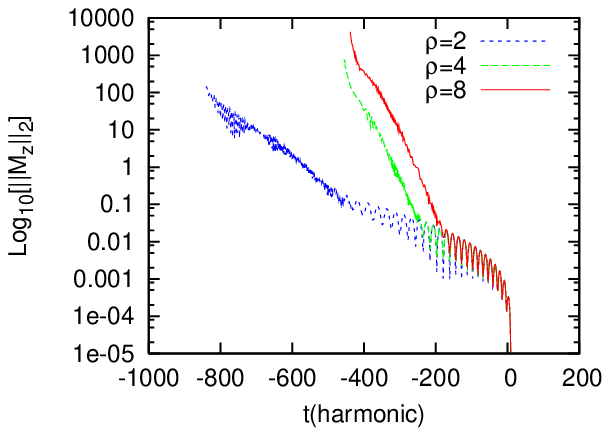}}\\
    \end{tabular}
    \caption{\label{1D-Gowdy-B2a}
     Collapsing polarized Gowdy-wave test with the adjusted BSSN system in 
     the $\tilde{\gamma}$-equation (\ref{B2a-adj}),  with 
     $\kappa_{\tilde{\gamma}}=0.000025$. 
     The figure style is the same as Figure~\ref{1D-Gowdy-plain}. 
     Note the almost 
     perfect overlap for 200 crossing-time in the both the Hamiltonian 
     and Momentum constraint and the $\rho=2$ run can evolve stably 
     for 1000 crossing-time.}
  \end{center}
\end{figure*}

%**************************************************************
The case of the adjustment of the $\tilde{\gamma}$-equation 
using the ${\cal G}$-constraint (\ref{B2a-adj}), is shown in 
Fig.~\ref{1D-Gowdy-B2a}. 
The adjustment parameter $\kappa_{\tilde{\gamma}}$ is 
set at $0.000025$.
(Again, the signature of $\kappa$ is reversed from the expected one.)

Figure~\ref{1D-Gowdy-B2a} shows that an almost 
perfect overlap is obtained for the 200 crossing-time in both ${\cal H}$ and ${\cal M}_z$. 
The higher resolution runs tend to crash at earlier times, which is same 
as with the plain BSSN system.
However, the convergence time becomes longer than that of the plain BSSN system. 
We will discuss the quantitative improvement for the 
$\tilde{\gamma}$-adjustment in the next subsection.

%-----------------------------------------------------------------------------
%-----------------------------------------------------------------------------

%********************************
\subsubsection{Adjustment effect}
\begin{figure*}
  \begin{center}
    \includegraphics[width=90mm]{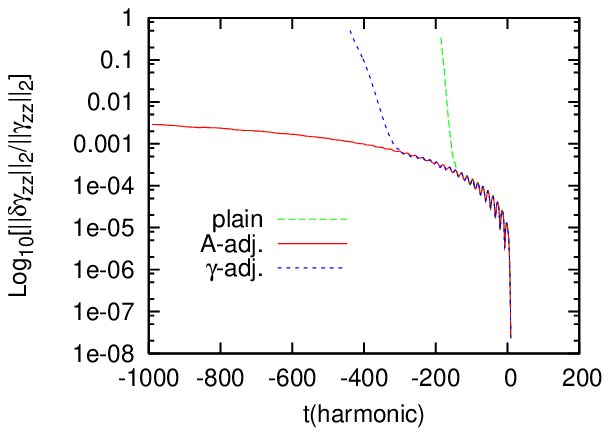}
    \caption{Comparisons of systems in the collapsing polarized Gowdy-wave test. 
     The L2 norm of the error in $\gamma_{zz}$,
     rescaled by the L2 norm of $\gamma_{zz}$, for the plain BSSN,
     adjusted BSSN with $\tilde{\cal A}$-equation, and with $\tilde{\gamma}$-equation are shown. 
     The highest resolution run, $\rho=8$, is depicted for the plots. We can conclude that 
     the adjustments make longer accurate runs available. 
     Note that the evolution is backwards in time.}\label{1D-Gowdy-gzz}
  \end{center}
\end{figure*}

%********************************
In order to check the {\it accuracy} of the simulations, we prepare 
Fig.~\ref{1D-Gowdy-gzz} to show 
the error of the $\gamma_{zz}$ component of the metric.

Unlike the gauge-wave or the linear wave test, in this Gowdy-wave test 
the amplitude of the metric functions damps with time. 
Therefore we use the criterion that the error normalized 
by $\gamma_{zz}$ be under $1\%$ for an {\em accurate evolution}.
 This criterion is the same as the one used in Zlochower et al.
~\cite{Zlochower:2005bj}. 

Figure~\ref{1D-Gowdy-gzz} shows the normalized 
error in $\gamma_{zz}$ versus time for the plain BSSN, 
adjusted BSSN with $\tilde{\cal A}$-equation, and 
adjusted BSSN with $\tilde{\gamma}$-equation systems. 
We find that these three systems 
produce accurate results up 
to $t=200$, $t=1000$, and $t=400$, respectively. 
This proves that the adjustments work effectively, i.e, 
they make possible a stable and accurate simulation, especially 
the ${\cal A}$-adjusted BSSN system.
%-----------------------------------------------------------------------------
%-----------------------------------------------------------------------------

%**************************************************
\section{Summary and Discussion}\label{sec:summary}
%**************************************************
In this article, we presented our numerical comparisons of the BSSN 
formulation and its adjusted versions using constraints.  
We performed three testbeds: gauge-wave, linear wave, and collapsing polarized
Gowdy-wave tests with their evolutions by 
three kinds of adjustments, which were previously proposed 
by Yoneda and Shinkai~\cite{Yoneda:2002kg} based on their constraint propagation analysis. 

The idea of the adjusted systems is to construct a system robust against
constraint violations by modifying the evolution equations  
using the constraint equations.

We can summarize our tests as follows:
\begin{itemize}
\item 
When the plain (original) BSSN evolutions
already show satisfactory good evolutions (e.g., the linear wave test), 
the constraint violations (i.e., adjusted terms) are also small or ignorable. 
\end{itemize}
Therefore the {adjusted} BSSN equations become quite similar to
the plain BSSN equations, and their results 
coincide with the plain BSSN results. 
\begin{itemize}
\item 
Among the adjustments we tried, we observed that 
the adjusted BSSN system with the $\tilde{A}$-eq. (\ref{B1-adj}) 
is the most robust for all the testbeds examined in this study. 
It gives us an accurate and stable evolution compared to the plain BSSN system. 
Quantitatively, the life-time of the simulation becomes 10 times longer 
for the gauge-wave testbed and 5 times longer for the 
Gowdy-wave testbed than the life-time of the plain BSSN system. 
However, it should be noted 
that for the gauge-wave testbed, the convergence feature is lost at 
a comparatively early time, the 200 crossing-time in the Hamiltonian constraint 
and the 50 crossing-time in the momentum constraint.  
\end{itemize}
Recently, it has been claimed that the set up of the gauge wave problem in 
Apples-with-Apples has a problematic point~\cite{Babiuc:2007vr}, 
which arises from the harmonic gauge condition. 
In \cite{Babiuc:2004pi}, it is argued that this gauge has a 
residual freedom in the form 
$H \to e^{{\lambda}t}H$, where $\lambda$ is an arbitrary and $H$ is 
a function in Eq.~(\ref{eq:gwave-metric}). 
Of course, 
our set up corresponds to the $\lambda=0$ case, but numerical 
error easily excites modes that result in either 
exponentially increasing or decaying metric amplitude. 
Actually, we find the amplitude of the error decays with time in 
this testbed. So, we conclude that 
due to the adjustment, the growing rate of the gauge mode is 
suppressed and the life-time of the simulation is extended as 
a result.
\begin{itemize}
\item 
The other type of adjustments (\ref{B2a-adj} and \ref{B2b-adj}) 
show their apparent effects while depending on a problem. 
The $\tilde{\Gamma}$-adjustment for the gauge-wave testbed makes 
the life-time longer slightly. 
The $\tilde{\gamma}$-adjustment for the Gowdy-wave testbed makes 
possible a simulation twice as long as 
the plain BSSN system. 
\end{itemize}
We can understand the effect of the adjustments in terms of 
adding dissipative terms.
By virtue of the definition of the constraints, we can recognize 
that the adjusted equation corresponds to the diffusion equation 
(see, for example, Eq.~(\ref{B1-adj})) and 
the signature of $\kappa$ determines whether the diffusion is 
positive or negative. 
In the adjusted $\tilde{A}$-eq. system, (\ref{B1-adj}), 
the adjustment term corresponds to 
the positive diffusive term, due to the definition of ${\cal M}_i$ 
and the positiveness of $\kappa_{A}$ (see Eq.~(\ref{eq:cal Mi}) 
and (\ref{B1-adj})). 
This fact might explain why the adjusted 
$\tilde{A}$-eq. system works effectively for all the testbeds. 

In contrast, why are not all the adjustments effective in all testbeds?
As we mentioned 
in Sec. IIB, the eigenvalue analysis was made on the linearly perturbed 
violation of constraints on the Minkowski space-time. Since the constraint 
violation grows non-linearly as seen in the Appendix of \cite{Yoneda:2002kg}, 
the candidates may not be the best in their later evolution phase.

\begin{figure*}
  \begin{center}
    \includegraphics[width=90mm]{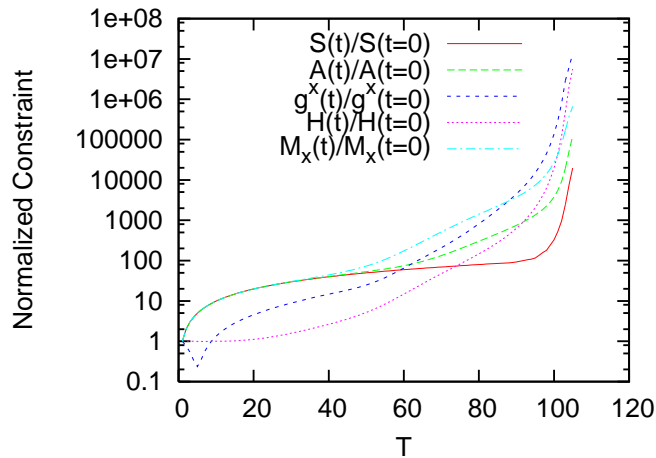}
    \caption{\label{fig:C-vio-Gauge}
    The violation of all constraints normalized with their initial values, 
    $||\delta{\cal C}||_2(t)/||\delta{\cal C}||_2(0)$, are plotted 
    with a function of time.  The evolutions of the gauge-wave testbeds
    with the plain BSSN system are shown.  
    The parameters of the test are the same as 
    those shown in Sec.~\ref{sec:Gaugewave}, and the lowest resolution run,
    $\rho=2$, is applied.
    By observing which constraint 
    triggers the other constraint's violation from the constraint propagation equations, 
    we may guess the mechanism by which the entire system 
    is violating accuracy and stability.  See the text for details.
}
  \end{center}
\end{figure*}

We remark upon two more interesting aspects arising 
from our study. 
The first is the mechanism of the constraint violations. 
As was shown in the appendix of~\cite{Yoneda:2002kg}, 
each constraint propagation (behavior of their growth or decrease) 
depends on the other
constraint terms together with itself. 
That is, we can guess ${\cal A}$ and ${\cal S}$ constraints 
(\ref{eq:cons-calA} and \ref{eq:cons-calS}) in this article, propagate 
independently of the other constraints, while the violation of the 
${\cal G}$-constraint, 
(\ref{eq:cons-calG}) is triggered by 
the violation of the momentum constraint, 
and both the Hamiltonian and the momentum constraints 
are affected by all the other constraints.  
Such an order of the constraint violation 
can be guessed in Fig.~\ref{fig:C-vio-Gauge} (earlier time), 
where we plot
the rate of constraint violation 
normalized with its initial value, 
    $||\delta{\cal C}||_2(t)/||\delta{\cal C}||_2(0)$,
as a function of time, for the gauge-wave testbeds
with the plain BSSN evolution. 
(Note that the constraints at the initial time, $\delta{\cal C}(0)$, are 
not zero due to the numerical truncation error. )
The parameters  are the same 
as those shown in Sec.~\ref{sec:Gaugewave}, and the lowest resolution 
run is used. 
From this investigation, we might conclude
that to monitor the momentum constraint violation is the key to checking the
stability of the evolution. 
%-----------------------------------------------------------------------------
%-----------------------------------------------------------------------------

The second remark is on the Lagrange multipliers, $\kappa$, used in the adjusted
systems. 
As discussed in Sec.~\ref{subsec:adjBSSN}, the signatures of the 
$\kappa$s are
determined {\it a priori}, and we confirmed that all the predicted 
signatures of $\kappa$s in~\cite{Yoneda:2002kg} are right to produce 
positive effects for controlling constraint violations.  
However, we have to search for a suitable magnitude of $\kappa$s 
for each problem.  
Therefore we are now trying to develop a more sophisticated
version, such as an auto-controlling $\kappa$ system, which will be reported 
upon in the future elsewhere. 

Although the testbeds used in this work are simple, 
it might be rather surprising to observe the expected effects of adjustments
with such a slight change in the evolution equations. 
We therefore think that 
our demonstrations imply a potential to construct a robust system 
against constraint violations even in highly dynamical situations, such as 
black hole formation via gravitational collapse, or binary merger problems. 
We are now preparing our strong-field tests of the adjusted BSSN systems using 
large amplitude gravitational waves, black hole space-time, or non-vacuum space-time, 
which will be reported on in the near future. 

%*************************
\section*{Acknowledgments}
%*************************
K.K. thanks K.~I. Maeda and S. Yamada 
for continuing encouragement. K.K. also 
thanks Y. Sekiguchi and M. Shibata for their 
useful comments on making numerical code. 
This work was supported in part by the Japan Society for 
Promotion of Science (JSPS) Research Fellowships 
and by a Grant-in-Aid for Scientific Programs.
H.S. was partially supported by the Special Research Fund 
(Project No. 4244) of the Osaka Institute of Technology. 
A part of the numerical calculations was carried out 
on the Altix3700 BX2 at YITP at Kyoto University.
%**************************

%********************

\end{document}